\documentclass[draftclsnofoot,12pt,onecolumn,web]{ieeetran}
\usepackage{cite}
\usepackage{amsmath,amssymb,amsfonts,gensymb}
\usepackage{algorithmic}
\usepackage{graphicx}
\usepackage{textcomp}
\usepackage{comment}
\usepackage{float}

\begin{document}
\title{Ferroelectric HfO$_2$ Memory Transistors with High-$\kappa$ Interfacial Layer and Write Endurance Exceeding $10^{10}$ Cycles}
\author{Ava Jiang Tan, Yu-Hung Liao, Li-Chen Wang, Jong-Ho Bae, Chenming Hu, Sayeef Salahuddin
\thanks{This work was supported in part
by the Berkeley Center for Negative Capacitance Transistors, the ASCENT Center, one of the six centers within the DARPA/SRC JUMP initiative and the DARPA FRANC program. The work of A. J. Tan was supported by the National Defense Science and Engineering Graduate Fellowship (NDSEG).}
\thanks{A. J. Tan, Y-H. Liao, L-C. Wang, C. Hu, and S. Salahuddin are
with the Department of Electrical Engineering and Computer Sciences, University
of California, Berkeley, Berkeley, CA 94720 USA (e-mail: ava@eecs.berkeley.edu).}
\thanks{J-H. Bae was with the University of California, Berkeley. He is
now with the Department of Electrical Engineering, Kookmin University, Seoul 02707,
South Korea.}}

\maketitle
\setcounter{page}{1}

{\begin{abstract}
  We demonstrate ferroelectric (FE) memory transistors on a crystalline silicon channel with endurance exceeding $10^{10}$ cycles. The ferroelectric transistors (FeFETs) incorporate a high-$\kappa$ interfacial layer (IL) of thermally grown silicon nitride (SiN$_x$) and a thin 4.5 nm layer of Zr-doped FE-HfO$_2$ on a $\sim$30 nm SOI channel. The device shows a $\sim$ 1V memory window in a DC sweep of just $\pm$ 2.5V, and can be programmed and erased with voltage pulses of $V_G= \pm$ 3V at a pulse width of 250 ns. The device also shows very good retention behavior. These results indicate that appropriate engineering of the IL layer could substantially improve FeFET device performance and reliability. 
\end{abstract}

\section{Introduction}
\label{sec:introduction}
Despite the fact that there has been much rekindled interest in ferroelectrics-based nonvolatile memories due to the discovery of CMOS-compatible doped HfO$_2$ materials \cite{reliability_DRC, pesic, mulaosmanovic,bae, chatterjee, liu, lee, sidoped_VLSITSA}, one of the key roadblocks facing the development of FeFETs specifically is endurance. It has been shown, for thicker $>$5-6 nm FE oxides, that bulk charge-trapping and interfacial
layer breakdown (due to the large coercive fields associated with ferroelectric
HfO$_2$, and therefore, the larger write voltages needed to program FeFETs with
thicker FE oxides) tend to cause premature device failure \cite{yurchuk,
mulaosmanovic2}, with many groups reporting typical endurance metrics of
$10^4$-$10^6$ cycles \cite{camcell_edl, chen, ali,florent2017first}. For thinner FE oxides $<$5nm in physical thickness, hot electron-induced hole damage and channel/oxide interface degradation tend to be the key agents limiting device endurance \cite{hote_VLSI}.

Proposals to extend the cycling lifetime of FeFETs include interfacial oxide
engineering, gate workfunction engineering, modulating the material properties
of the FE layer itself, and many others \cite{nitrided_EDL, deng, ni,
chernikova, cao, oh}. 
%
%
In this work, we combine a high-$\kappa$ interfacial layer (IL) together with a thin FE film ($\sim$4.5 nm). This is motivated by the previous observation by many authors (such as \cite{chatterjee}) that the endurance cycling is mostly limited by IL breakdown. In fact, in a metal-FE-metal capacitor configuration, cycling endurance metrics exceeding 10$^{10}$ are routinely observed \cite{pesic}. In a recent report, Abhishek et. al. circumvented the interfacial layer breakdown problem by fabricating a bottom-gate, channel last transistor, where an oxide semiconductor channel was grown directly on the FE material, thus achieving an endurance exceeding 10$^{12}$ cycles \cite{abhishekintel}. Similarly, Kim et. al. recently reported on the fabrication of vertical 3D NAND FE thin film transistors utilizing indium zinc oxide as the semiconductor channel, showing a cycling endurance of up to $10^8$ \cite{kim}. Nonetheless, when crystalline Si is used as the channel material, as is required for high performance memory, formation of an IL is inevitable, and therefore endurance still remains a critical challenge to be addressed.

In the context of IL breakdown, it is known that `time-to-breakdown' has an exponential relationship to the applied electric field in the interfacial layer \cite{schuegraf1994hole}. In other words, a mild decrease in the electric field could still lead to a substantial increase in the `time-to-breakdown', and therefore could slow the generation of traps that eventually counteract the FE hysteresis. For the same charge density, a high-$\kappa$ IL reduces the electric field by the ratio of its permittivity to that of SiO$_2$. Our choice of high-$\kappa$ IL is thermal nitridation of chemically formed silicon oxide. This provides a simple way to achieve an IL with a permittivity $\sim 8$. Thermally grown silicon nitride also has a comparable breakdown field to SiO$_2$\cite{ito1978thermally}. A thin 4.5 nm of FE HZO is chosen to suppress the effects of bulk charge trapping \cite{hote_VLSI} and demonstrates the thickness scalability of FE HZO. We show that this combination substantially improves the device performance. In a DC sweep, almost a 1V memory window can be achieved with just $\pm 2.5$V. More importantly, with bipolar stress pulsing at $\pm 3$V, 250 ns, the endurance exceeds 10$^{10}$ cycles on silicon.

\section{Experiment and EOT Comparison}
\label{sec:experiment}

The structure of the FeFET device characterized in this work is shown in Fig.
\ref{fig1}(a), and TEMs to compare its gate stack incorporating a nitrided IL
against that of a baseline device with an SiO$_2$ IL (characterized in previous
work \cite{hote_VLSI}) are shown in Fig. \ref{fig1}(c) and Fig. \ref{fig1}(b),
respectively. The process flow to realize the FeFET is described in
\cite{camcell_edl}, with the IL formation step involving thermal nitridation of
the SOI substrate at 850 $\degree$C in NH$_3$ ambient rather than a
self-terminated chemical growth of SiO$_2$. As confirmed through TEM, the FE
oxide thickness of both the control SiO$_2$ FeFET and the FeFET with a nitrided
IL are the same (roughly 4.5 nm after 45 cycles of deposition). The IL
thicknesses of the SiO$_2$ IL and nitrided IL are $\sim$8$\AA$ and $\sim$1.5 nm,
respectively.

Fig. \ref{fig1}(d) compares the CV of the baseline FeFET with a SiO$_2$ IL to
the CV of the FeFET with a nitrided IL. Though the physical thickness of the
gate stack of the latter is larger, its capacitance is also larger. Using Synopsys TCAD, we have estimated the net EOT of the nitrided sample to be roughly 1 $\AA$ smaller than the baseline sample, based upon the accumulation capacitances. This allows us to make an estimate for the effective $\kappa$ of the IL as follows:

\begin{equation}
    \kappa_{\text{NIL}}= \kappa_{\text{SiO}_2}\times \frac{t_{\textrm{NIL}}}{t_\textrm{baseline}-\delta \textrm{EOT}_{\textrm{net}}}=3.9\times\frac{15}{7.5}=7.8 \nonumber
\end{equation}
\\
\noindent where $\kappa_\text{NIL}$ and $\kappa_{\text{SiO}_2}$ indicate the $\kappa$ values of the nitrided IL and SiO$_2$ IL respectively; $t_\text{NIL}$ and $t_{\text{baseline}}$ indicate the physical thicknesses of the nitrided IL and SiO$_2$ IL respectively; and $\delta\text{EOT}_{\text{net}}$ is the simulated EOT difference between the two ILs. This calculation indicates the IL is nearly all Si$_3$N$_4$. Therefore, we expect to reduce the electric field in the IL layer by two times, which will ultimately result in a substantial increase in the time to breakdown. 

\section{Results and Discussion}
\label{sec:resultsanddisc}

We first investigate the DC hysteresis of the fabricated device with a nitrided IL. Fig. \ref{fig2}(a) shows results of a doubly swept $I_DV_G$ curve. Nearly a 1V memory window can be achieved with just $\pm 2.5$V sweep. We note that, compared to published literature, this is quite a low voltage requirement. For example, our baseline devices as reported in \cite{hote_VLSI} do not demonstrate any appreciable memory window at $\pm 2.5$V. Nonetheless it is also well known that the time to switch a given amount of polarization depends strongly on the applied voltage. Therefore, although the DC sweep is a good way of visualizing the hysteresis, it is important to also probe the high-speed switching behavior. Fig. \ref{fig2}(b) and (c) show measured current at a read voltage of $V_G=\pm  0.25$V as a function of pulse width. We define the high current/low $V_T$ state as the \texttt{ERS} state, and the low current/high $V_T$ state as the \texttt{PGM} state. Unsurprisingly, we observe a strong dependence of the current with the applied voltage. Below 1 $\mu$s, 2.5V is not good enough to provide the current level we see in the DC hysteresis. As the voltage amplitude increases, the current increases, signifying switching of a larger amount of polarization. At $V_G=3$V, the current approaches the level seen in DC hysteresis, even for a pulse width of $\sim$ 100 ns. Similarly, for the \texttt{PGM} state, $V_G=-3$V brings the current level down to almost the level seen in the DC hysteresis at a pulse width of $\sim$ 250 ns. The asymmetry between \texttt{PGM} and \texttt{ERS} states is expected -- accumulation of a thin SOI body requires a much larger voltage drop across the semiconductor, as discussed in previous reports\cite{bae}. Therefore, for a reasonably fast and symmetric operation, we choose a pulse width of 250 ns and a gate voltage of $V_G=\pm 3$V for endurance cycling.

In many studies, the endurance is quantified by measuring $\pm V_T$ after a certain number of bipolar stress pulses. The $\pm V_T$ determination requires one to perform sweeps over a small voltage range, which typically takes $\sim 1$ second to complete. On the other hand, the importance of fast reading has recently been discussed (e.g. \cite{abhishekintel}). It is known from charge pumping experiments (as discussed in \cite{pesic2, hote_VLSI}) that beyond several $\mu$s, charge trapping/de-trapping starts to manifest. These effects could in principle be quite complex, and could arise from the interplay between traps with varying time constants. Therefore, while slow sweeps to determine $\pm V_T$ could provide important insights into trap assisted phenomena, they are also expected to \emph{artificially} affect the actual currents that will be observed in a real application where the device is read quickly. Due to these considerations, we adopt fast reading of the device to determine its state. The complete endurance testing protocol is detailed in Fig. \ref{endurance_test}(a). During the stressing phase of the endurance test, bipolar voltage pulses of $\pm3$V, 250 ns are applied at the gate of the FeFET, with a 250 ns delay between sequential pulses to achieve a stressing period of 1 $\mu$s total in duration. Periodically, a state determination test is conducted to evaluate the margin between the \texttt{PGM} and \texttt{ERS} states. For this, a 10 $\mu$s read pulse is applied at the gate of the FeFET (after either the \texttt{ERS} or \texttt{PGM} pulse), after ramping and stabilizing the drain voltage to 50 mV (see Fig. \ref{endurance_test}(a), right panel). The averaged current value during this 10 $\mu$s reading period is determined to be the read current. Figs. \ref{endurance_test}(b) and (c) show the readout current during the 10 $\mu$s reading period for the high $V_T$ (\texttt{PGM}) and low $V_T$ (\texttt{ERS}) states, respectively. In both cases, the current saturates well within the READ duration of 10 $\mu$s.

Fig. \ref{endurance_char}(a) shows the results of endurance testing across different devices, proving that the devices with a nitrided IL can be reliably cycled to $10^{10}$. The data is plotted as $I_{\text{ERS}}/I_{\text{PGM}}$ vs. fatigue cycles for ease of comparison. The exact current levels are shown for a device in Fig. \ref{endurance_char}(b). First, we note that the high and low current levels are very similar to those measured from a DC sweep. This indicates that the device is switched properly with the \texttt{PGM/ERS} pulses. Interestingly, the device does not show any rapid degradation after $10^4-10^6$ cycles, as reported in most studies. Rather, the high current level shows a slow degradation. The envelope of the low current similarly shows a slow degradation (increase); yet the separation of the current levels retains a margin of $10^3$  until $6 \times 10^{10}$ cycles. Beyond that point, a sudden breakdown is observed. Notably, this sudden breakdown is correlated with the gate leakage through the device shooting up several orders of magnitude, as shown in Fig. \ref{endurance_char}(c), indicating that the gate oxide itself breaks down close to $10^{11}$ cycles. We note that, there are devices that does not experience such a breakdown. For example,  Fig. \ref{endurance_char}(d) shows DC $I_DV_G$ sweep from an exemplary device cycled to $10^{12}$. The anti-clockwise Hysteresis is still clearly visible. This should compared with previous studies where total reversal of handedness of Hysteresis happens after just 10$^4$-10$^5$ cycles. This also shows that the Ferroelectric itself remains quite robust even after one trillion cycles.  

Finally, we present the retention behavior. As seen in Fig. \ref{retention}, the retention looks unaffected for a testing duration of $10^4$ seconds for both the \texttt{PGM} and \texttt{ERS} states. The retention looks robust at both 25$^\degree$C and 85$^\degree$C. Thus despite relatively lower voltage operation and very large endurance, there is no discernible effect on the retention behavior.

\section{Conclusion}
\label{sec:conclusion}
 In conclusion, we have demonstrated a FeFET memory device with an engineered high-$\kappa$ IL that shows larger than 10$^{10}$ endurance cycles at a relatively small \texttt{PGM/ERS} voltage of $V_G=\pm 3$V and pulse width of 250 ns. Endurance measured on multiple devices show robust and repeatable behavior over 10$^{10}$ endurance cycles. We have identified total  oxide  breakdown  as  the  main limiting  factor as opposed to defect (interface and bulk) induced clock-wise Hysteresis that has been reported by many previous studies.   Understanding charge injection  during the  endurance test and optimizing to reduce the total oxide breakdown could increase the endurance over 10$^{12}$ cycles, as our data shows that the FE film remains robust even beyond that cycling number.  Additional optimization of the IL and FE layers could allow for further reduction in the operating voltage to below 2V, while maintaining and/or even enhancing the endurance behavior. One potential drawback of using such a high-$\kappa$ IL could be a reduction in mobility. However, somewhat reduced mobility could still be tolerable as these devices are not expected to compete with the gate-delay of logic devices.

}
\newpage 
{\begin{figure}[H]
\centerline{\includegraphics[width=0.9\columnwidth]{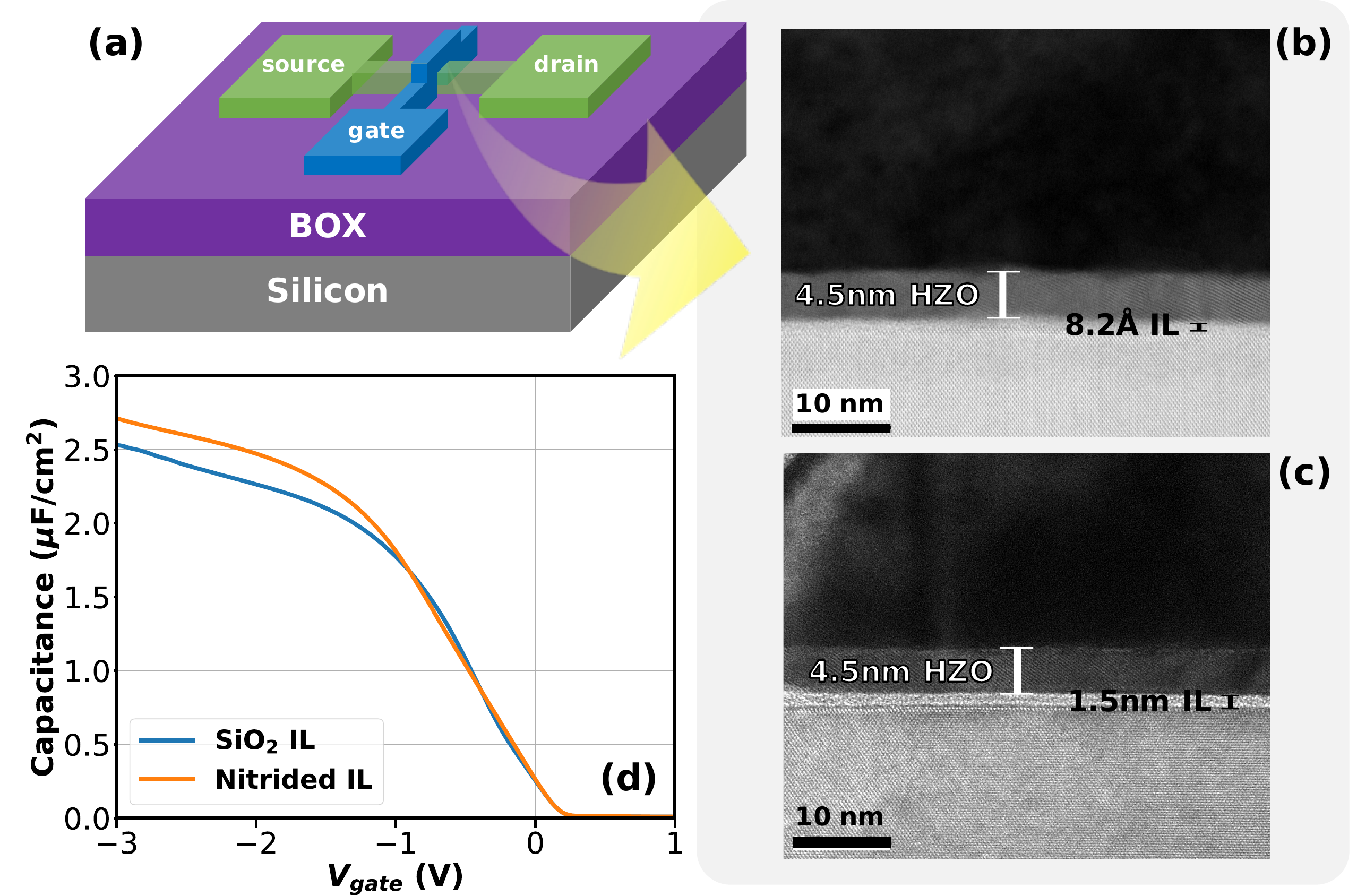}}
  \caption{(a) Schematic of a SOI, gate-first FeFET device. (b) TEM of
  4.5 nm HZO  gate stack with 8.2$\AA$ SiO$_2$ IL. (c) TEM of 4.5 nm HZO gate
  stack with 1.5 nm SiN$_\text{x}$IL. (d) CV comparison of gate stack with SiO$_2$ IL
  to the gate stack with nitrided IL, both taken at 100 kHz.}
  \label{fig1}
\end{figure}

\begin{figure}[H]
\centerline{\includegraphics[width=0.9\columnwidth]{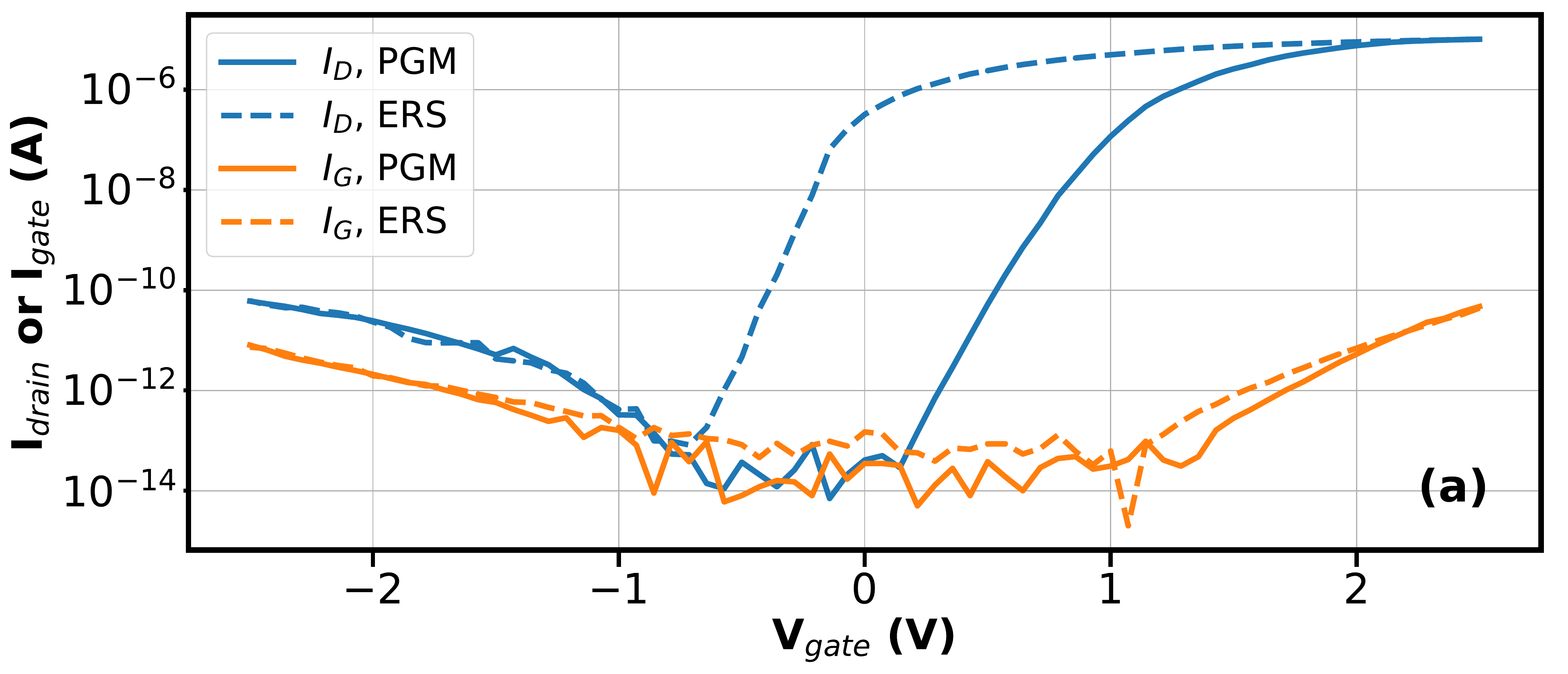}}
\centerline{
   \includegraphics[width=0.45\columnwidth]{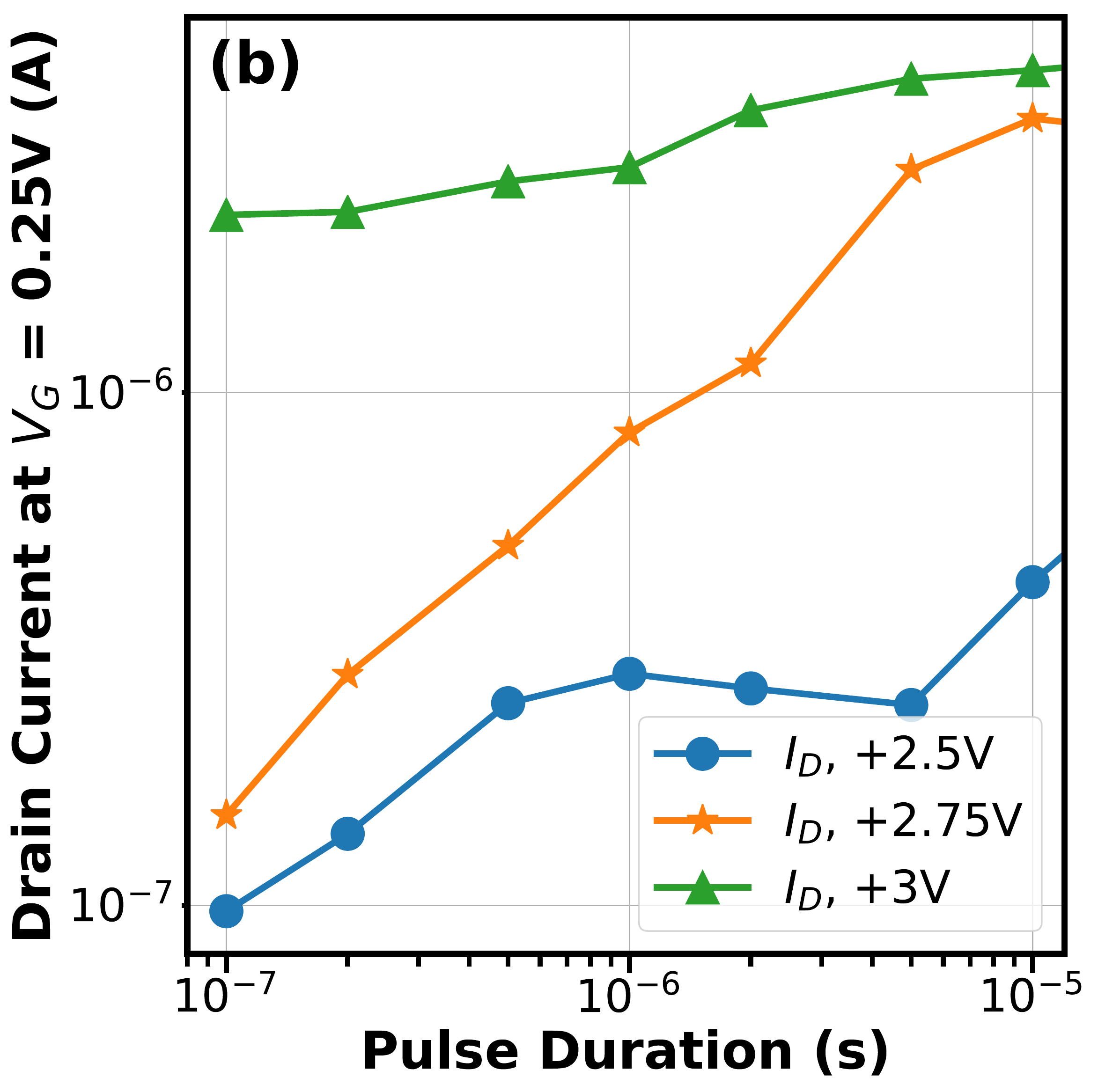}
   \includegraphics[width=0.45\columnwidth]{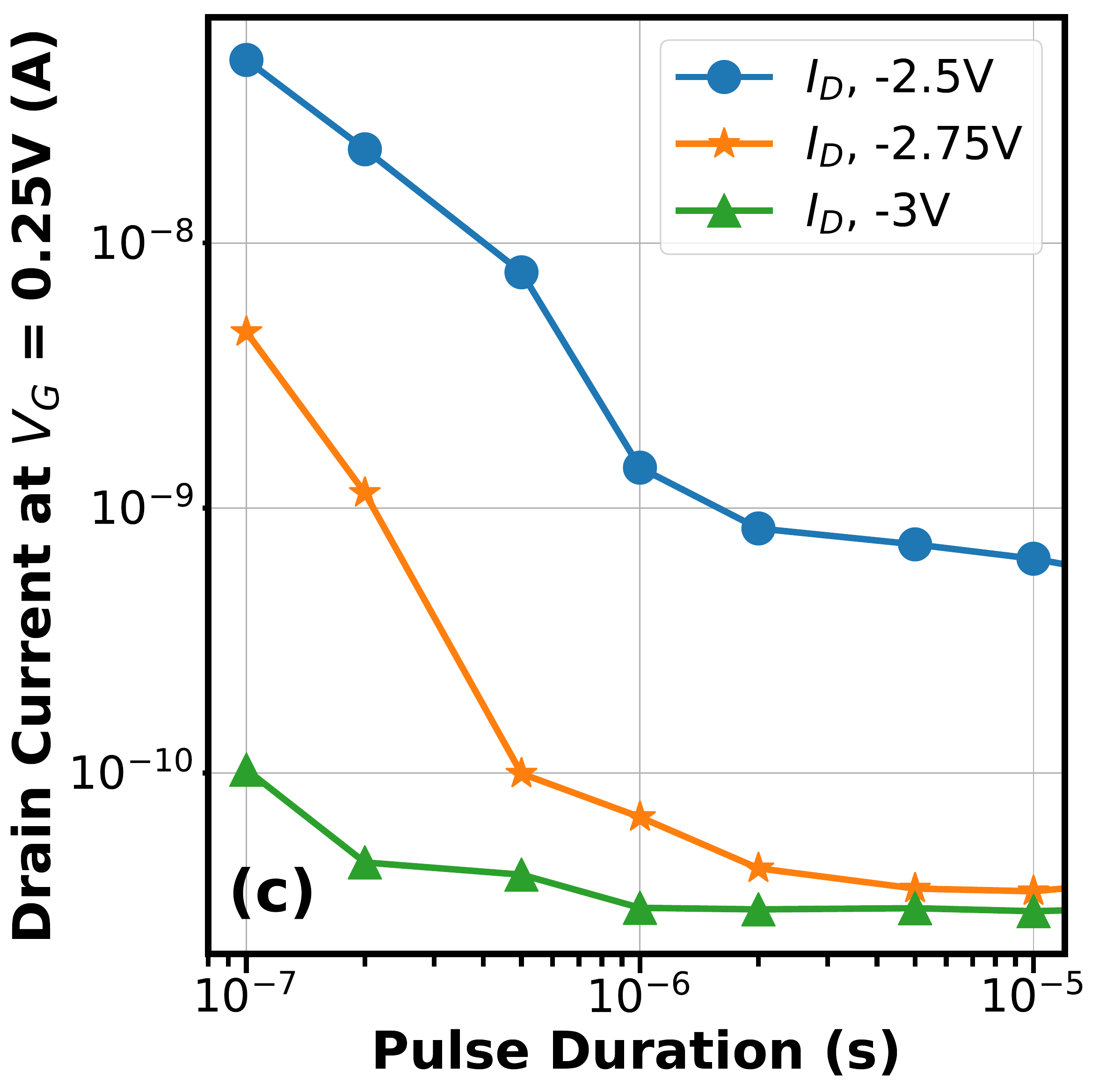}}
 \caption{(a) $I_DV_G$ of a typical FeFET with 4.5 nm HZO on a 1.5 nm nitrided IL. The device is doubly-swept from $\pm 2.5$V at a drain bias of $V_D = $50 mV. (b) Typical \texttt{ERS} characteristics for the FeFET. (c) Typical \texttt{PGM} characteristics for the FeFET. Voltage magnitudes range from $\pm 2.5 - 3$V and pulse durations from 100 ns to 10 $\mu$s.}
\label{fig2}
\end{figure}

\begin{figure}[H]
\centerline{\includegraphics[width=0.9\columnwidth]{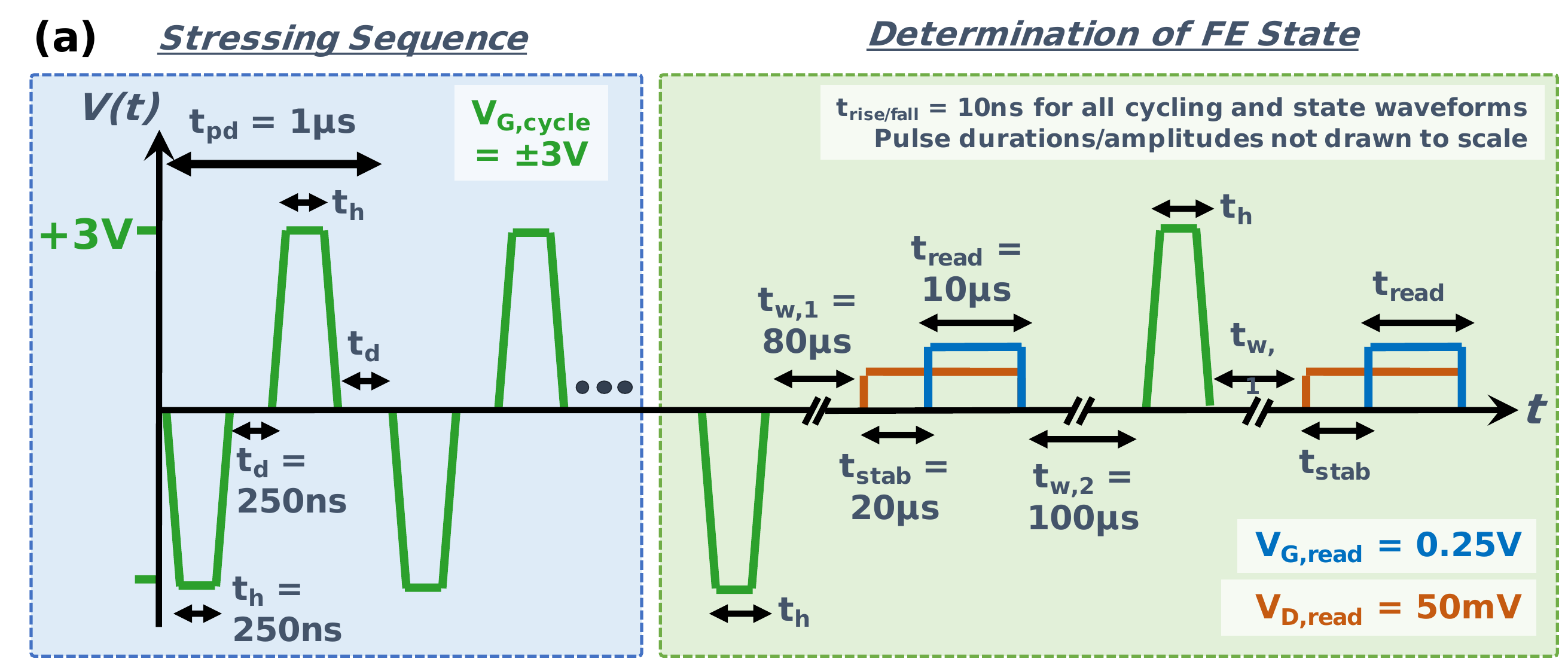}}
\centerline{\includegraphics[width=0.45\columnwidth]{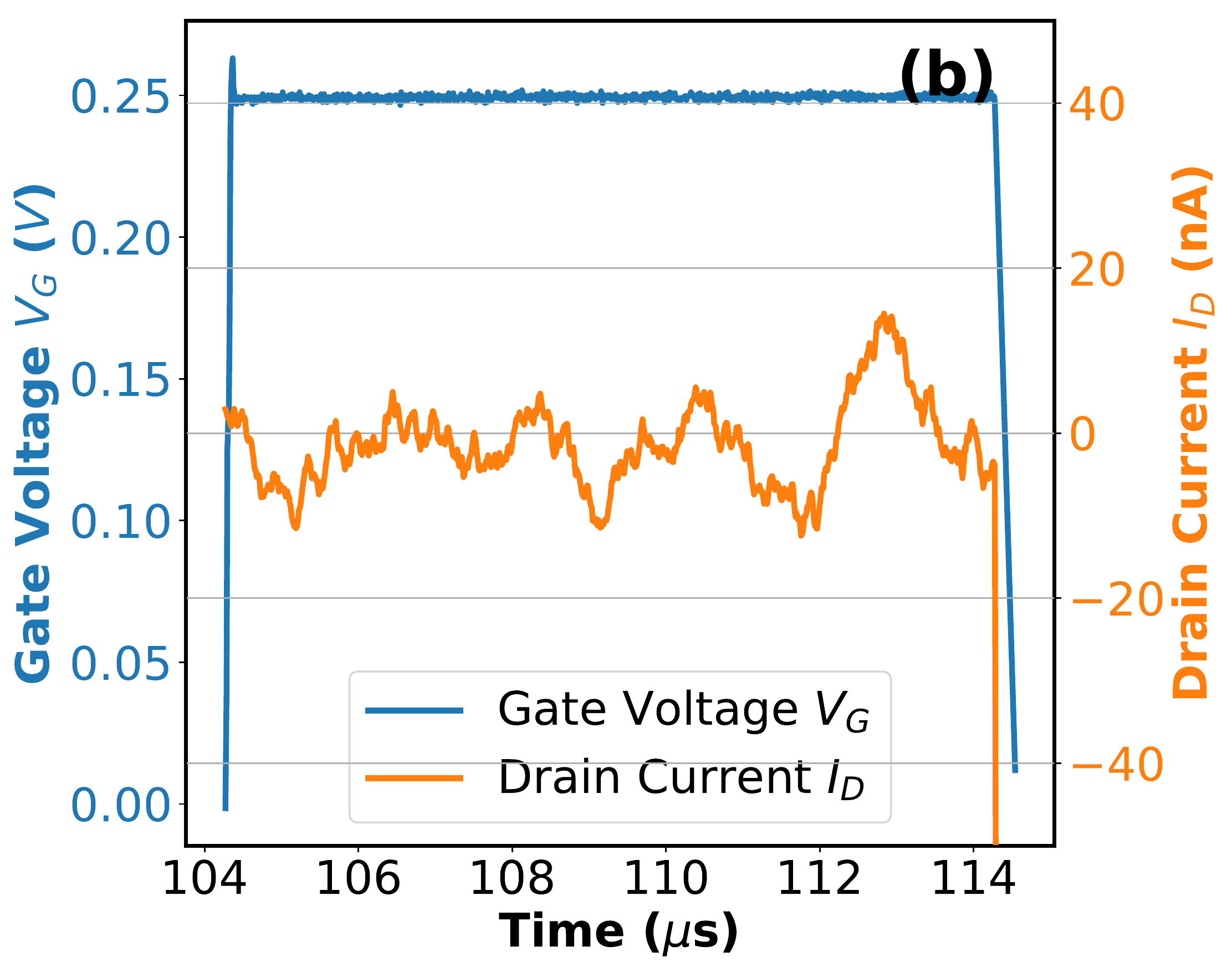}\includegraphics[width=0.45\columnwidth]{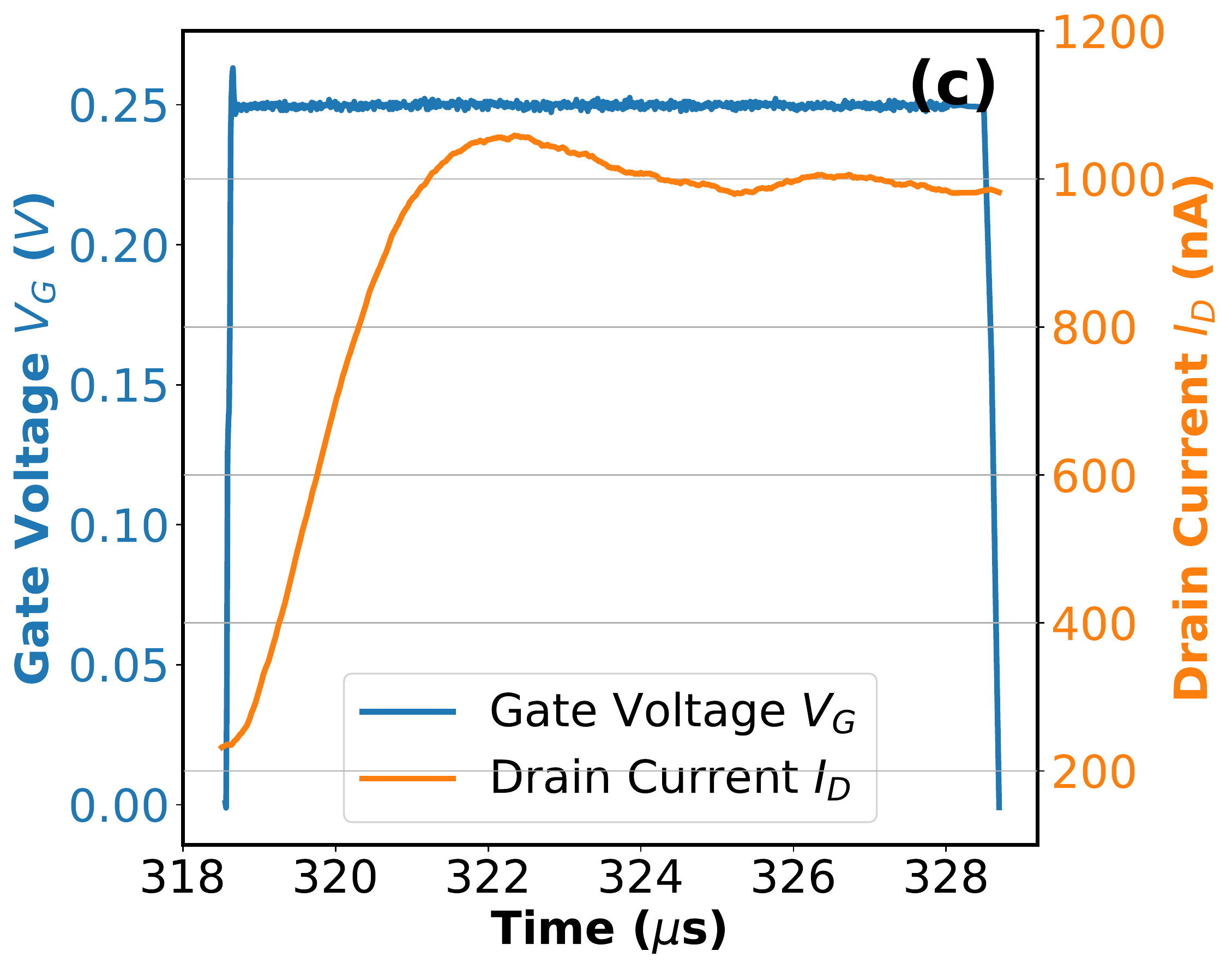}}
  \caption{(a) Endurance stressing sequence and subsequent ferroelectric state
  determination waveforms used to characterize the FeFETs in this work. (b) Transient current readout waveform corresponding to the high $V_T$ state (and low readout current). (c) Transient current readout waveform corresponding to the low $V_T$ state (and high readout current).}
\label{endurance_test}
\end{figure}

\begin{figure}[H]
\centerline{\includegraphics[width=0.9\columnwidth]{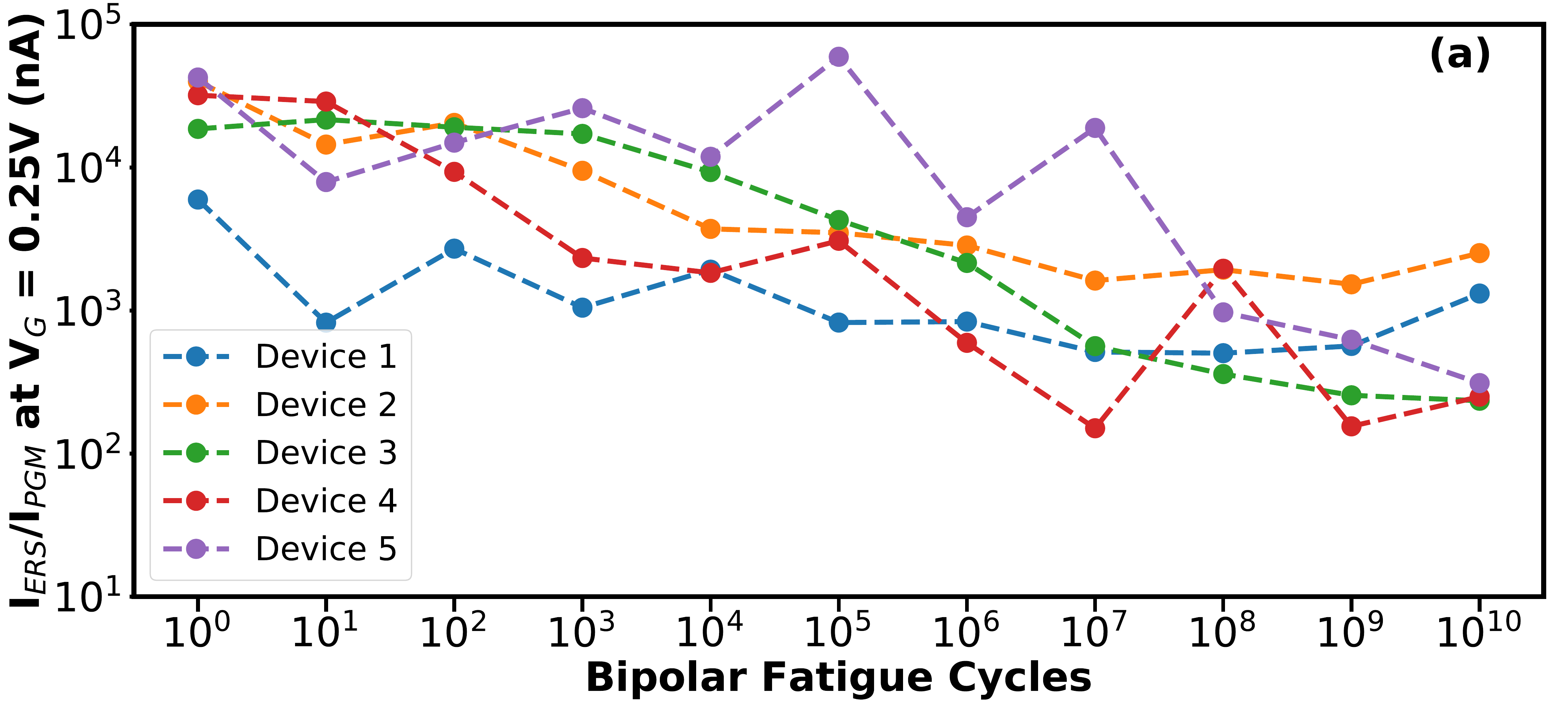}}
\centerline{\includegraphics[width=0.9\columnwidth]{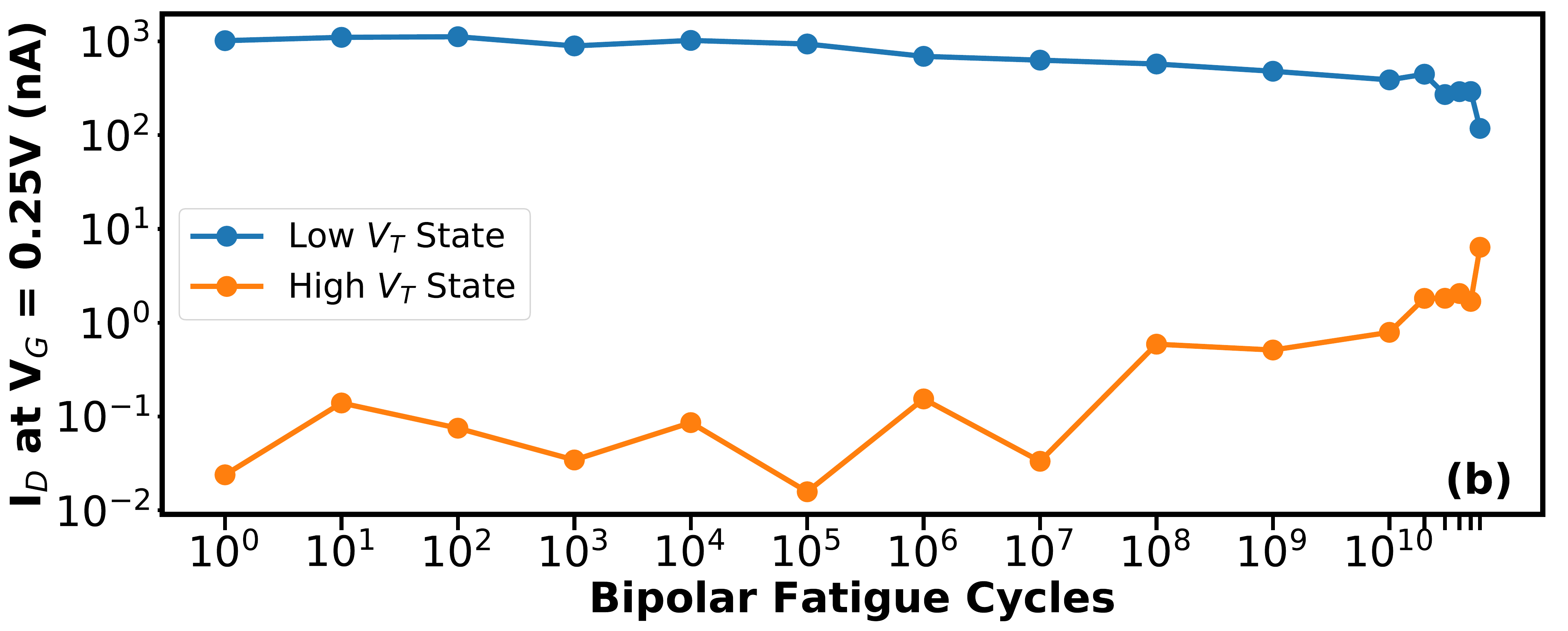}}
\centerline{\includegraphics[width=0.35\columnwidth]{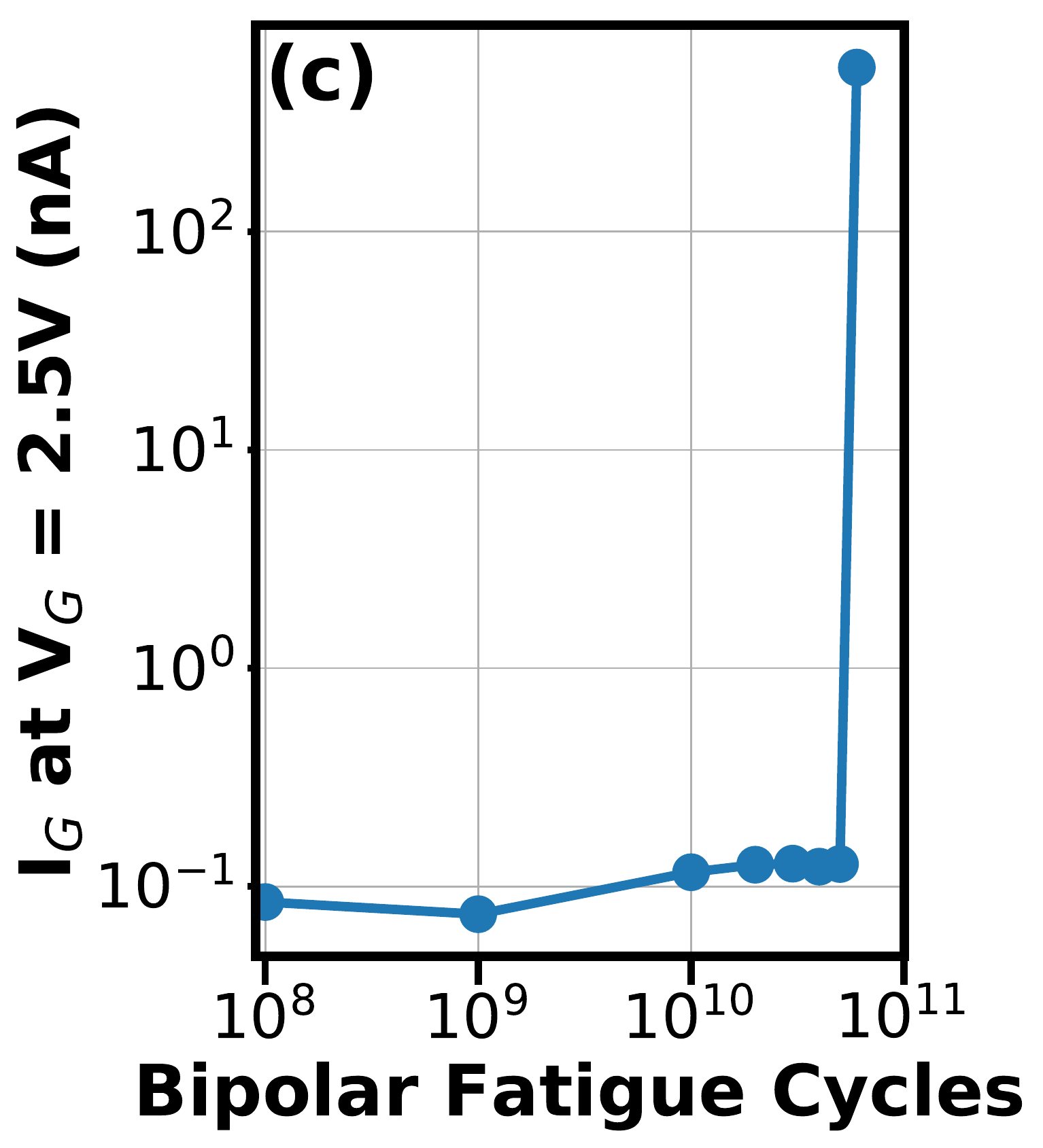}\includegraphics[width=0.52\columnwidth]{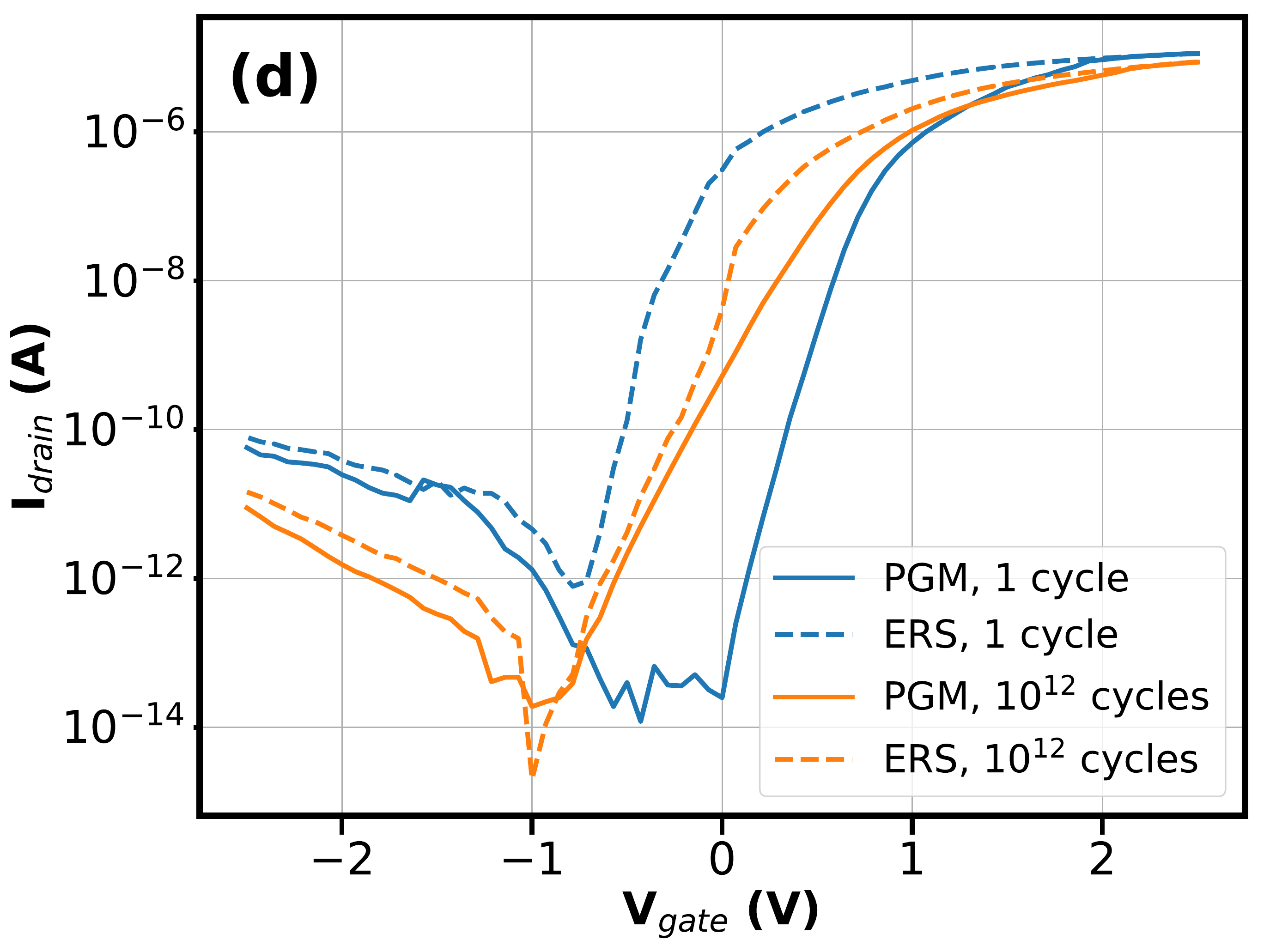}}
  \caption{(a) Endurance characteristics of multiple FeFET devices cycled from 1 to $10^{10}$ fatigue cycles. Results are reported as the ratio of the low $V_T$ $I_D$ readout to the high $V_T$ $I_D$ readout at every decade of cycling. (b) A device cycled until $\sim$1 order of magnitude of current separation remains after $6\times10^{10}$ cycles. (c) $I_G$ at $V_G =2.5$V as a function of cycling for the same device in (b), showing a strong correlation between oxide wearout and loss of memory window. (d) $I_DV_G$'s of an exemplary device cycled to $10^{12}$ cycles, showing some remaining ferroelectric hysteresis at the end of the endurance test.}
\label{endurance_char}
\end{figure}

\begin{figure}[H]
  \centerline{\includegraphics[width=0.7\columnwidth]{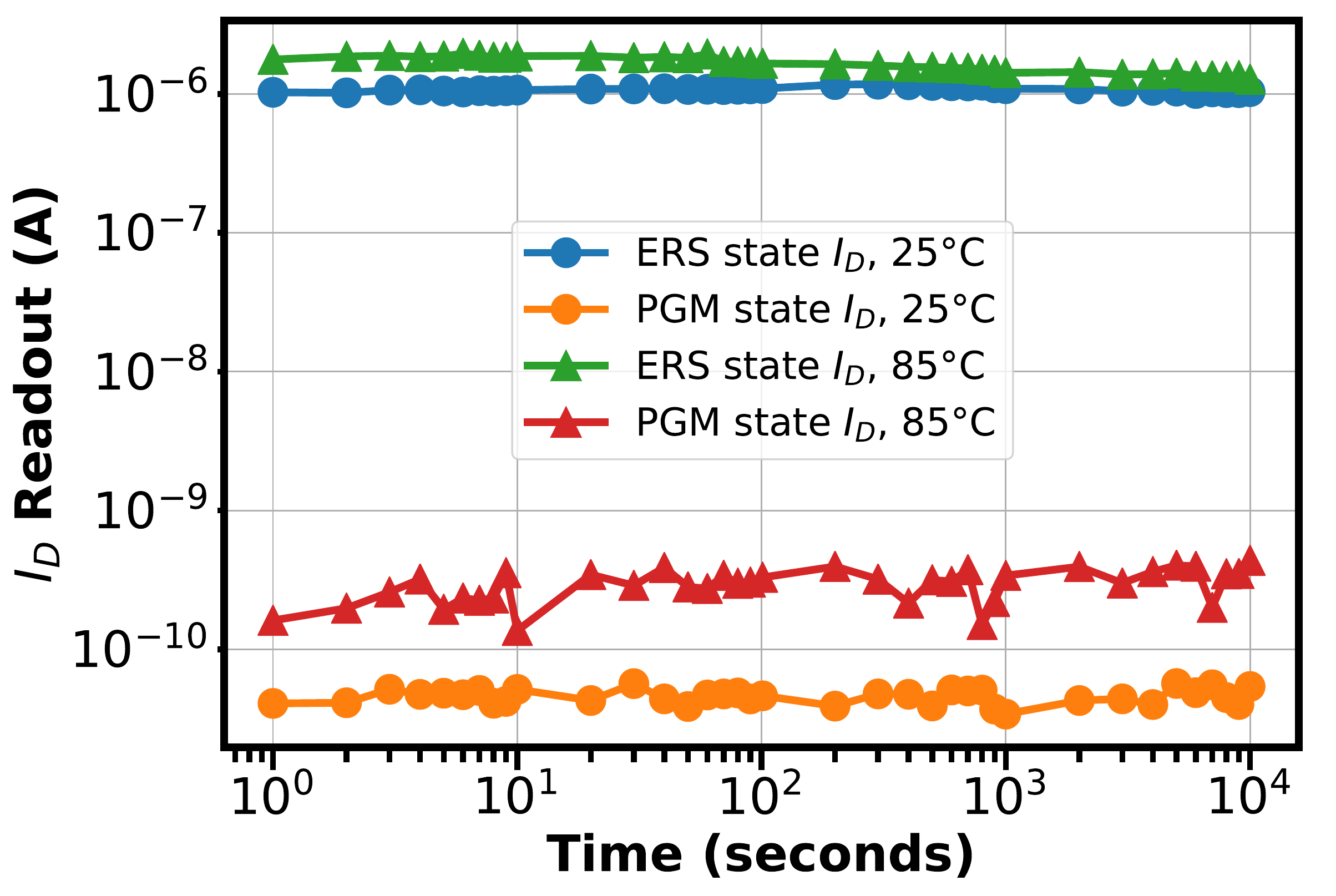}}
\caption{Retention testing at room temperature (25 $\degree$C) and at elevated
  (85 $\degree$C) for $10^4$ seconds. Gate read voltage is chosen to the same at both testing conditions after correcting for the leftward $V_T$ shift due to an effective substrate doping change at elevated temperature.}
\label{retention}
\end{figure}
}
\newpage
\bibliographystyle{IEEEtranDOI}
\bibliography{references}
\end{document}